\documentclass[reprint,showpacs,amsmath,amssymb,aps]{revtex4-1}

\usepackage{graphicx}
\usepackage{dcolumn}
\usepackage{bm}

\begin{document}

\title{Results of wavelet processing of the $2K$-capture $^{78}$Kr experiment statistics}

\author{Yu.M.~Gavrilyuk}
\author{A.M.~Gangapshev}%
\author{V.V.~Kazalov}%
\author{V.V.~Kuzminov}%
\affiliation{Baksan Neutrino Observatory INR RAS, Russia
}%

\author{S.I.~Panasenko}
\author{S.S.~Ratkevich}
\email{ratkevich@univer.kharkov.ua}
\affiliation{V.N.Karazin Kharkiv National University, Ukraine
}%


\begin{abstract}
Results of a search for $^{78}$Kr double $K$-capture with the large low-background proportional counter (2005-2008 years) at the Baksan
Neutrino Observatory are presented. An experimental method and characteristics of detectors are described.
Basic features of the digitized pulses processing using wavelet transform are considered. With due account taken of the analysis of
individual noise characteristic it has been shown that the appropriate choice of both wavelet
characteristics and sequence of processing algorithms allows one to decrease the background in the
energy region of useful events with a unique set of characteristics by $\sim2000$ times.
New limit on the half-life of $^{78}$Kr with regard to $2K$-capture has been found: T$_{1/2}\geq2.4\cdot10^{21}$
yrs (90\% C.L.).
\end{abstract}

\pacs{23.40.-s, 27.50.+e, 29.40.Cs, 98.70.Vc}

\maketitle

\section{\label{Intr}Introduction}
Most $\beta\beta$ decay investigations have concentrated on the $\beta^-\beta^-$ transition (two-neutrino emitting and neutrinoless modes). This transition for the two-neutrino mode has been discovered in direct and geochemical experiments for ten nuclei of the $^{48}$Ca, $^{76}$Ge, $^{82}$Se, $^{96}$Zr, $^{100}$Mo, $^{116}$Cd, $^{128}$Te, $^{130}$Te, $^{150}$Nd  and $^{238}$U \cite{Zdesenko},\cite{Barabash06}. The data obtained for $2\nu$-mode offer a chance to directly compare different models of the nuclear structure, which form the basis for calculations of
nuclear matrix elements $|M^{2\nu}|$, and to select the optimal one. Though direct correlation between the values of nuclear matrix
elements for the two-neutrino and neutrinoless modes of $\beta\beta$ decay is absent lacking, the methods for calculating
$|M^{2\nu}|$ and $|M^{0\nu}|$ are very close, and a chance possibility to estimate their accuracy in calculating $|M^{0\nu}|$
appears only when comparing experimental data and theoretical results calculations for the probability of $2\nu\beta\beta$ decay.

It can be expected that acquisition of experimental data on the other types of $\beta\beta$ transitions
[the decay with emission of two positrons $(\beta^+\beta^+)$, capture of bound atomic electrons with emission of a positron $(\beta^+EC)$ and double capture of two bound atomic electrons $(ECEC)$]
will make it possible to considerably increase the quality of calculations for both $2\nu$ and $0\nu$ processes.
Much efforts have been currently made in searching for these processes \cite{Danevich05},\cite{Barabash09} and \cite{Rukhadze09},
in spite of the fact that the $\beta^+\beta^+$ and $\beta^+EC$ modes are strongly suppressed relative to $\beta^-\beta^-$ decay
due to the Coulomb barrier for positrons, and a substantially lower kinetic energy attainable in such transitions. Positrons are
absent in the final state of the $2\nu ECEC$ transition,
and the kinetic energy of the transition may be rather high (up to 2.8 MeV), which
dictates determines an increased probability of a decay. However, this process is also difficult to detect, since it is
only characteristic radiation that is detectable in it.

The experiment to search for $2K$-caputre in $^{78}$Kr using a copper low-background proportional counter of
large volume has been carried out in the Baksan Neutrino Observatory INR RAS since June 2005 \cite{r1}.
The data obtained during first 159 hours of measurements with enriched isotope of krypton allowed us
to estimate the sensitivity ($S$) of the new installation in the experiment
to search for $^{78}$Kr $2K$-capture half-life.
In one year of measurements it has reached $ S = 1 \cdot 10^{22} $ year (90\% C.L.).

Theoretical calculations based on different models give the following half-life times for this process
in $^{78}$Kr: $3.7\cdot10^{21}$ years \cite{r2}; $4.7\cdot10^{22}$ years \cite{r3}; $7.9\cdot10^{23}$ years \cite{r4}.
The last two values were obtained from the estimation of the half-life time in $^{78}$Kr with regard to the
total number of $2\nu ECEC$-transition where 78.6\% of events is due to $2\nu2K$-capture \cite{r5}.
Comparison of experimental and theoretical values shows that the sensitivity of the measurements has
exceeded the lower limit of theoretical calculations.
This fact allows us to test the model of Aunola and Suchonen
\cite{r2}.
The technique to search for $2K$-capture
used in our research is based on the following considerations discussed below.

\section{\label{Basic}Basic assumptions}

When two electrons
are captured from the $K$-shell in $^{78}$Kr, a daughter atom of $^{78}$Se$^{**}$ is formed with two
vacancies in the $K$-shell.
The technique to search for this reaction is based on the assumption that the energies of
characteristic photons and the probability that they will be emitted when a double vacancy is filled are
the same as the respective values of the case when two single vacancies of the \emph{K}-shell in two singly
ionized Se$^*$ atoms are filled. In such a case, the total registered energy is $2K_{ab} = 25.3$ keV, where $K_{ab}$
is the binding energy of a \emph{K}-electron in a Se atom (12.65 keV). The fluorescence yield upon filling a
single vacancy of the \emph{K}-shell in Se is 0.596. The energies and relative intensities of the characteristic
lines in the \emph{K}-series are $K_{\alpha1} = 11.22$ keV (100\%), $K_{\alpha2} = 11.18$ keV (52\%), $K_{\beta1} = 12.49$ keV (21\%),
and $K_{\beta2}=12.65$ keV (1\%) \cite{x-ray}. The probability of deexcitation of a doubly ionized \emph{K}-shell through the emission of  Auger electrons $(e_a,e_a)$ only, or through a single characteristic quantum and an Auger electron $(K,e_a)$, or through two characteristic X-rays and low-energy Auger electrons $(K,K,e_a)$ is $p_1=0.163$, $p_2=0.482$ and $p_3=0.355$, respectively.

A characteristic photon can pass a long distance in the gas from the point of its origin to the point of its absorption. For example, 10\%
of characteristic photons of 12.6 keV energy are  absorbed on a length of 2.42 mm ($P_{gas}\sim4.36$ Bar) \cite{Storm73}.
Auger electrons of the same energy will be absorbed on a length of 0.44 mm, which
produce almost pointwise charge clusters of primary ionization in the gas \cite{Drift and diffusion}.
In the case of emission of two characteristic photons absorbed in a gas the energy release will be distributed between three pointwise regions. It is these events that have a number of unique features and are the subject of study in this paper.

\section{\label{technique}Experimental technique}

To register this process a large proportional counter (LPC) with a casing of M1-grade copper has been used.
LPC is a cylinder with inner and outer diameters of 140
and 150 mm.
A gold-plated tungsten wire of 10 $\mu$m in diameter is stretched along the LPC axis and serves as an anode.
The potential of +2400 V is applied to the wire,  and the casing (the cathode) is grounded.
Both ends of the
anode are electrically connected to the to the corresponding end cap flanges via high-voltage pressure-sealed ceramic insulators
with a central electrode taken from spark plugs.

To reduce the
influence of the edges
on the operating characteristics of the counter, the end segments of the wire are passed through the copper
tubes (3 mm in diameter and of 38.5 mm length) electrically connected to the anode. Gas amplification is
absent on these segments, and charge is collected in an ionization mode.
Taking into account teflon insulator dimensions, the distance from the operating region to the flange is 70 mm.

The length of the LPC operating volume is 595 mm
(the distance between the butt ends of the tubes),
and the LPC operating volume is 9.159 l. The total capacitance of the counter and outlet insulator is $\sim30.6$ pF. The total resistance of the anode and two output electrodes is $\sim600$ Ohm.
Indium wire is used to seal all detachable joints, and teflon gaskets are used to seal all nipple joints.
The inner insulators are made of teflon, and their
thickness was minimized to improve the degassing
conditions during the vacuum treatment of the counter and to stabilize its operating characteristics.

The LPC was placed inside the shielding of 18 cm thick copper, 15 cm thick lead, and 8 cm thick borated polyethylene layers.
The installation is located
in one of the chambers of the underground laboratory of the Gallium Germanium Neutrino Telescope experiment at the Baksan Neutrino Observatory, INR RAS, at a depth of 4700 m.w.e. where cosmic ray flux is lowered by  $\sim10^7$ times down to the level of $(3.03\pm0.10)\times10^{-9}$ cm$^{-2}$s$^{-1}$ \cite{r9}.

The counter is filled
with pure krypton gas up to 4.42 Bar having no quenching or accelerating gaseous
additions and is purified through a Ni/SiO$_2$ absorber from electronegative
admixtures.
Two samples of krypton
were used in this work: one of 48.6 l volume enriched in $^{78}$Kr up to 99.8\% \cite{Kr-78purf} and natural krypton (100 l) left after $^{78}$Kr isotope extraction.
Both samples have been specially cleared from cosmogeneous radioactive isotope
$^{85}$Kr (T$_{1/2}=10.756$ yrs), present in atmospheric krypton.

The detector signals are passed from the anode wire to the charge-sensitive amplifier (CSA) directly attached to the high-voltage insulator.
The CSA parameters have been optimized to transmit a signal with
minimum distortions, and information on the primary-ionization charge spatial
distribution in its projection to the counter radius is fully represented in the
pulse shape. After amplification in an auxiliary amplifier the pulses are collected
by the LA-n20-12PCI digital oscilloscope.
The oscilloscope, integrated with personal computer, records the pulse waveform as the numerical vector digitized with a frequency of 6.25 MHz.
The length of the scanning frame is 1024 points (163.8 $\mu$s) with a resolution of 160 ns, $\sim50$ $\mu$s is "prehistory"
and $\sim114$ $\mu$s  is "history".

Processing of digitized pulses in "offline" mode is carried out using the
technique developed in this work.
This technique rejects pulses
of non-ionzied nature, determines the type of event and its relative coordinate along the anode wire, and finally selects useful events according to chosen characteristics.

\section{\label{impulse}IDENTIFICATION OF MULTIPOINT EVENTS}

The response of a linear time-invariant system to
some influence can be expressed by the convolution integral:
\begin{equation}
 \label{Eq1}
 q(t) = \int\limits_{ - \infty }^\infty  {h(\tau )i(t - \tau )} d\tau,
\end{equation}
where in our case $i(t)$
is the electric current at the
 output of LPC, $q(t)$ is the signal from CSA,
$h(t)$ - the pulse transitive characteristic of CSA. To describe the output signal recorded by the digital
oscilloscope in the form of a discreet set of instantaneous values
of amplitudes one can move from integration to summation of instantaneous function values with a step  $\Delta t$:
\begin{equation}\label{Eq2}
q(k\Delta t) = \Delta t\sum\limits_n {h(n\Delta t)i(k\Delta t - n\Delta t)} ,
\end{equation}

The pulse shape of registered charged particles (or photons) taken from LPC has some definite features which allow us to easily
distinguish them from non-ionized noise signals and microbreakdown events.
The registered events, in their turn, can be sorted out as
"point-like" and "lengthy" events by comparing their pulse rise time.
Point-like pulses with a small rise time for the pulse front are produced by particles which interact
with a krypton gas in the counter
in some particular region whose characteristic length is small in comparison with the cathode radius.
Lengthy events with long of pulse rise time can be produced when several point-like events are simultaneously registered at different
distance from the anode ("multi-point" or "multi-cluster" events).
Besides, "extended" pulses arise in a case when the fast electron track length is commensurable with counter radius.

\begin{figure}[pt]
\includegraphics*[width=2.15in,angle=270.]{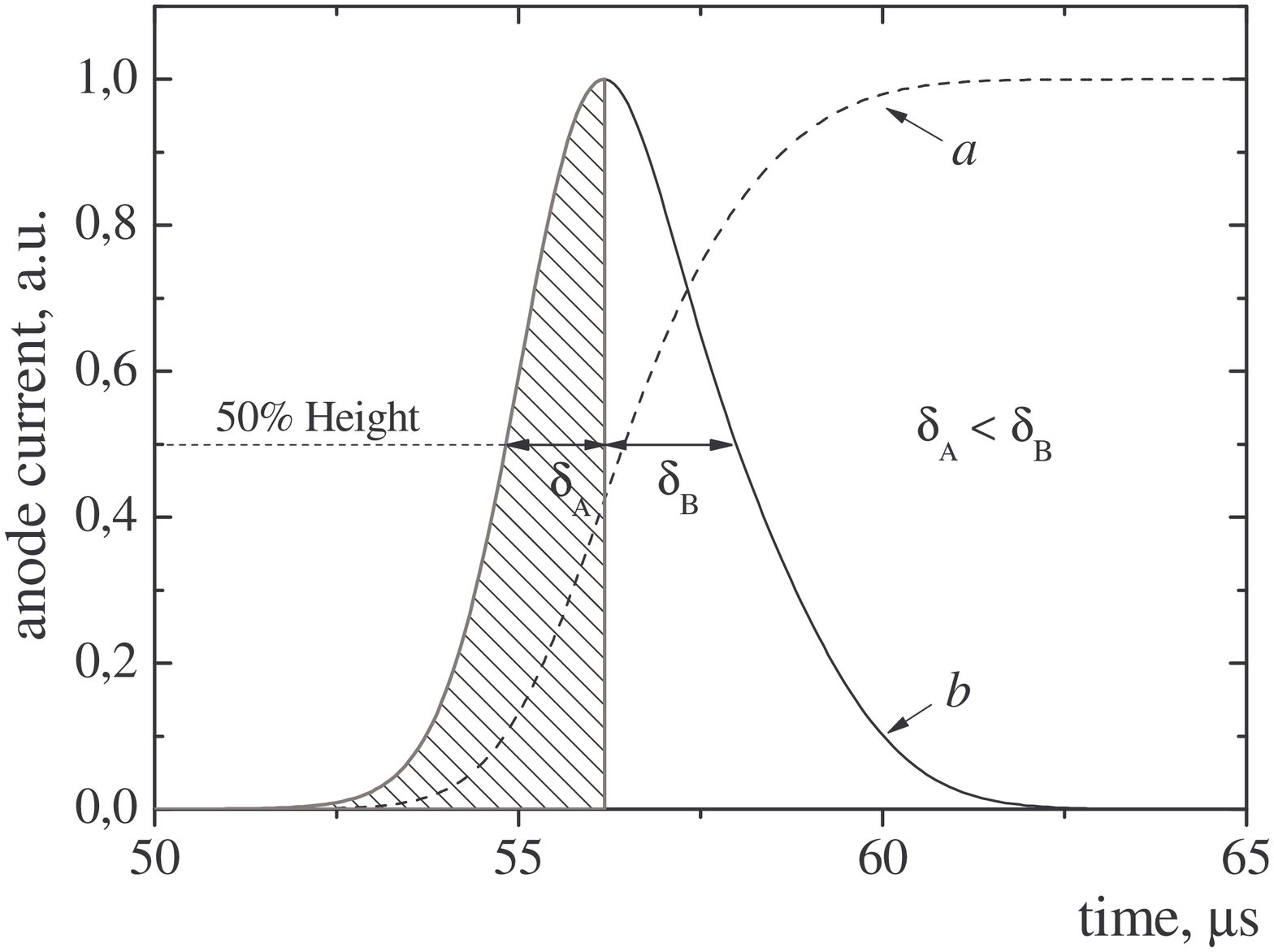}%
\caption{\label{fig:Lopsided Gausian} A
depiction of idealized pulse shape single-point event from LPC.
The monotonically dashed curve \emph{a} is a charge pulse
pulse from the CSA output,
and the front-back asymmetry solid curve \emph{b} is
is an electric current pulse
(constructed by differentiating the charge pulse).}
\end{figure}

Idealized pulse shape events taken at the output
of the LPC's and charge pulses taken from CSA's output  produced by single-point events are shown in Fig.~\ref{fig:Lopsided Gausian}.
The derivative of a pulse taken from CSA replicates the shape of a electric current signal from the LPC's anode wire. In a simplified case one can calculate the value of current in LPC by:
\begin{equation}\label{Eq3}
\overline i _n  = \frac{{\overline q _n  - \overline q _{(n - 1)} }}{{\Delta x}},
\end{equation}
where $\Delta x=x_n-x_{n-1}$.

A pulse taken from the anode wire is formed mainly due to the negative charge induced onto the anode by positive ions moving towards the cathode
and produced near the wire during the gas amplification
(ionic component - i.c.).
The contribution of the charge induced at the anode by electrons from avalanches
(electronic component - e.c.)  is small enough because they have to pass a small potential difference on their way to the anode.
The electric current pulse shape is affected
by the distribution in density of primary ionization electrons crossing the border of gas amplification region. The distribution parameters depend on drift time of the originally point-like charge towards the anode. The point-like ionization is smeared out during the drift time due to diffusion of the electrons into the cloud whose charge density distribution in its projection onto the radius is close to the Gaussian one.

\begin{figure}[pt]
\includegraphics*[width=2.25in,angle=270.]{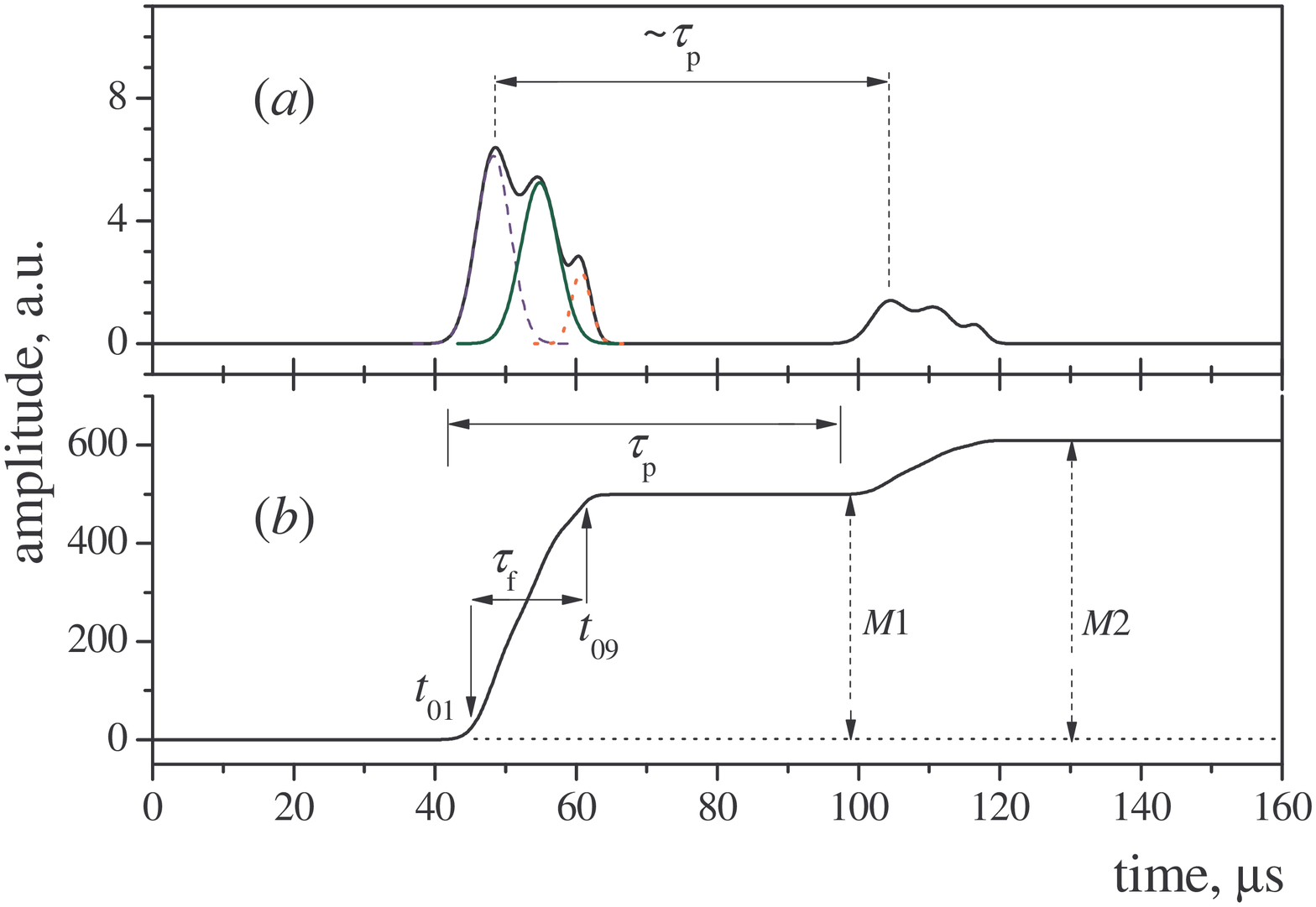}%
\caption{\label{fig:model_pulse} A depiction of idealized pulse shape analysis parameters from three-point event expected from $2K$-capture in case of registering two characteristic photon and a cascade of low energy Auger electrons from LPC: ({\it a})- the electric current  pulse, ({\it b})- the charge pulse.}
\end{figure}

As is seen in Fig.~\ref{fig:Lopsided Gausian}, the electric current signal produced as a result of gas amplification has an asymmetric form and cannot be described by a single Gaussian curve. There is a mathematical procedure which allows conversion from a registered charge pulse to a electric current pulse produced by primary ionization electrons through the conventional border of gas amplification.
The detailed description of the algorithm for electronic component segregation
is given in \cite{PTE}.
The obtained form of a signal can be described by a set of Gaussian curves, thus
determining the charge that was deposited in separate components of a multipoint
event.
The calculated area of an individual Gaussian should correspond to the charge (energy) of the corresponding point-like ionization.
Fig.~\ref{fig:model_pulse} illustrates a demo version of a shape of a electric current pulse $i(t)$: ({\it a}) from LPC and the corresponding CSA pulse ({\it b}) from three-point event expected from $2K$-capture in case of registering two characteristic X-rays and a cascade of low energy Auger electrons.
In Fig.~\ref{fig:model_pulse}, in addition to the primary pulse, there is a secondary pulse, smaller in amplitude, due to the photoelectrons knocked out of the copper case by photons produced in the electronic avalanches in the gas amplification process.

Photoeffect on the cathode is probable enough because of absence of quenching additions in the krypton gas. The delay between pulse and after-pulse is determined by the total drift time of the electrons to move from the cathode to the anode. It presets the duration of the time interval to allocate totally any single event regardless of  its primary ionization distribution over the LPC volume. In case of pure krypton the calculated drift time for the ionization electrons to move from the cathode to the anode is 58 ${\mathrm \mu}$s.  A Gaussian area or a sum of Gaussians in case of a multipoint event for an interval of 58 $\mathrm{\mu} s$, starting from the beginning of a pulse, gives total number of primary ionization electrons.

For the pulse shape analysis we have used the following signal parameters indicated in Fig.~\ref{fig:model_pulse}:
1) $\tau_\mathrm{f}=t_{09}-t_{01}$ is the pulse rise time; $t_{01}$ and $t_{09}$  are the moments where the pulse amplitude has
reached 10\% and 90\% of its maximum value $M1$; 2) $\tau_\mathrm{p}$  is the time of the after pulse appearance since the beginning of the primary pulse; 3)  $\lambda=(M2-M1)/M1$ is the ratio of the difference between the after-pulse's and primary pulse's maximum amplitudes to the primary pulse amplitude.

The parameters borders corresponding regions where useful signals are expected, and are determined from the distribution plotted for the pulses
with known characteristics obtained
from the calibration sources \cite{PTE}.
Using these borders
we select the candidate-events for $2K$-capture from the whole set of events registered in the course of basic measurements.

\begin{figure}[pt]
\includegraphics*[width=2.25in,angle=270.]{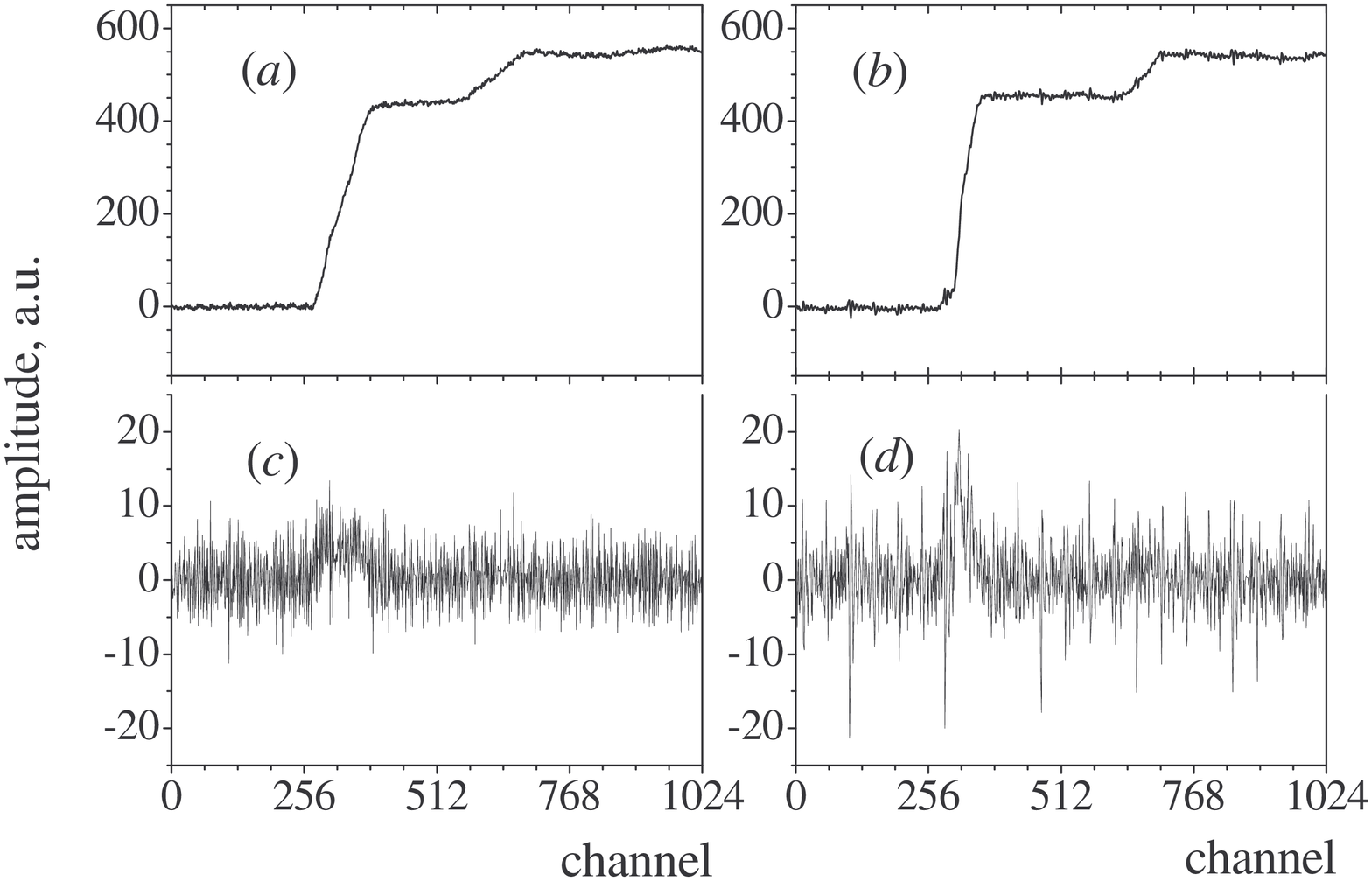}%
\caption{\label{fig:real_pulse} $(a)$ and $(b)$
curves demonstrate the converted
pulses of voltage (charge) normalized for $M1$ of an original CSA pulse and obtained by integrating electric current pulses due to primary ionization electrons $(c)$ and $(d)$ from two real events which are candidates for 2K-capture of $^{78}$Kr.}
\end{figure}

Fig.~\ref{fig:real_pulse} $(a)$ and $(b)$ show converted
pulses of voltage (charge) normalized for $M1$ of an original CSA pulse and obtained by integrating electric current pulses due to primary ionization electrons [$(c)$ and $(d)$] from two real events which are candidates for $2K$-capture of $^{78}$Kr.

As is seen from Fig.~\ref{fig:real_pulse} $(c)$ and $(d)$ the electric current  pulses are very noisy. Noise and
possible electrically-induced signals can both mask the low-energy component and produce a false one. Smoothing procedures applied to the initial pulse, with a following differentiation of it, result in deterioration of electric current pulse resolution. As to traditional frequency filtering methods with various window functions like filters of Hemming, Winner, Sawitsky-Goley and others \cite{PTE} applied to the electric current signal, it does not
allow one to always separate with enough reliability closely located other components of a compound event masking each other.

\subsection{
Usifeul signal extraction from noisy LPC electric current pulses with wavelet analysis}
At present the mathematical base for wavelet analysis is well developed and present a good alternative to the Fourier transform in studying
time (spatial) series with distinct irregularity.
Wavelet transform, for a one-dimensional signal in particular, consists in its decomposition, through zooming and transfers,
on a basis constructed from a wave-like oscillation function with definite properties called wavelet.
Each function of this basis is characterized both by a definite spatial (time) frequency and its localization in the physical space (time).
The localized basis functions of wavelet transform resemble the signals under investigation to a greater degree than the Fourier basis functions.

Thus, the distinction between the Fourier transform, conventionally used for signal analysis, and the wavelet transform is that the latter
provides two-dimensional decoupling of a regular signal where frequency and coordinate are considered to be independent variables,
which means that in the wavelet transform an increase in resolution of one of these variables does not automatically result in the decrease of resolution of the other.
This independency of both variables allows one to simultaneously analyze the properties of a signal  both in physical (time, coordinate) and frequency spaces \cite{Daub92}-\cite{Wavelets}.
The wavelet transform has also a distinction from the short-time Fourier transform (the Gabor transform) resulting in a two-dimensional spectrum.
The distinction is that one and the same function constitutes the basis of the former while in the case of the latter the functions are different \cite{Daub92}, and thus the wavelet transform provides the better time-frequency localization, flexibility of analysis and possibility to chose the best fitting form of the wavelet function.

For our applications, the extraction of a real signal from noisy data is better approach mathematical tool of discrete wavelet transform (DWT).
DWT algorithm is related to based upon uses the discreet vector $\{f(n)|n\in N\}$, for which a wavelet spectrum is caluculated which in its turn is also a discreet vector.
For the orthonormal wavelet bases on a binary lattice, there have been developed algorithms of multiresolution analysis (MRA) or multiscale approximation (MSA). The idea of the latter consists in representing a signal by a sequence of images with different degree of detailing \cite{Mallat89}. Each image contains independent  nonoverlapping information about a signal in the form of wavelet coefficients which are easily calculated by the iteration procedure known as Fast Wavelet Transform.  Taken as a whole, they solve the problem of the total analysis of a signal and substantially simplify diagnostics of the underlying processes.

MRA application allows one to represent the signal under investigation, $\tilde{x}(n)$ , in the form of decomposition:
\begin{equation*}
x^J (n) = \sum\limits_{m = 0}^{N_{j_o }  - 1} {\tilde{a}_{j_0 ,m} } \varphi _{j_0 ,m} (n) + \sum\limits_{j = j_0 }^J {\sum\limits_{m = 0}^{N_j  - 1} {\tilde{d}_{j,m} \psi _{j,m} } } (n) ,
\end{equation*}
where $\varphi _{j_{\rm{o}},m} (n) = \sqrt {2^{j_{\rm{o}} } } \varphi (2^{j_{\rm{o}} } n - m)$ and $\psi _{j,m}(n) = \sqrt {2^j } \psi (2^j n - m)$
are the known  orthonormal scaling functions $\varphi _{j_{\rm{o}},m} (n)$ and wavelet functions $\psi _{j,m} (n)$, respectively ;
$\tilde{a}_{j_0 ,m}  = \left\langle {\tilde{x},\varphi _{j_o ,m} } \right\rangle $ are empirical approximation coefficients and
$\tilde{d}_{j,m}  = \left\langle {\tilde{x},\psi _{j,m} } \right\rangle$ are empirical detail coefficients;
$j,m \in  Z$ are the current  values of scale and shift;
$N_{j_o}$ $(N_j$) is the number of approximation coefficients (detail coefficients) considered at the corresponding levels of decomposition;
$j_o$ is the initial scale value; $J$ is the final scale value.

The resolution of the wavelet reconstruction $x^J (n)$ of an initial signal  $x(n)$ is set by parameter $J$.
Eventually $ \parallel {x^J  - \tilde{x}} \parallel _{_2} \rightarrow 0$ with  $J \rightarrow \infty$.

Scaling functions $\{ \varphi_{j,m}(n) \}$ and mother wavelet functions $ \{ {\psi _{j,m}(n)} \}_{j,m \in Z} $
with $2\xi$ non-zero coefficients satisfy the so-called two-level relations \cite{Daub92}:
\begin{eqnarray*}
\varphi (n) &=& \sqrt 2 \sum\limits_{k = 0}^{2\xi - 1} {h(k)\varphi (2n - k)}, \\
\psi (n) &=& \sqrt 2 \sum\limits_{k = 0}^{2\xi - 1} {g(k)\varphi (2n - k)},
\end{eqnarray*}
where $h(k)$ and $g(k)$ are the coefficients of low and high frequency filters of wavelet transform where
$g(k) = \left( { - 1} \right)^k h(2\xi - k - 1)$.

The discrete orthogonal wavelet-transform at MRA can be defined as a recursive algorithm:
\begin{eqnarray*}
a_{(j-1),m}&=&\sum\limits_{k \in N} {h_{k-2m}} a_{jk}, \\
d_{(j-1),m}&=& \sum\limits_{k \in N} (-1)^k g_{-k-2m+1} a_{jk}, \\
a_0 &=&\tilde{x}.
\end{eqnarray*}
According to the Mallat pyramid algorithm \cite{Mallat99},  the initial signal is first passed through the decomposition
filters of low and high frequency.  The next step is to obtain approximation coefficients ${\tilde{a}_{j,m}}$
at the output of low frequency filter and detail coefficients ${\tilde{d}_{j,m}}$ at the output of high
frequency filter, using decimation $\downarrow2$. This algorithm goes on according to the scheme given below
until it reaches the decomposition level $J$. In terms of of wavelet coefficients the wavelet decomposition of
a signal is presented in the following way.
\begin{eqnarray*}
W(\tilde{x}) = \tilde{a}_0 \rightarrow \{\tilde{a}_1, \tilde{d}_1  \}
\rightarrow \{\tilde{a}_2,\tilde{d}_2, \tilde{d}_1  \}  \rightarrow ...\\ ...\rightarrow \{\tilde{a}_N,
\tilde{d}_N , \tilde{d}_{N-1},...,\tilde{d}_1  \}
\end{eqnarray*}

The solution of the noise reduction problem is carried out in four steps:
1) the initial signal is decomposed;
2) the threshold value for noise is chosen for each level of decomposition;
3) the threshold filtration of detail coefficients is carried out;
4) the signal is reconstructed.

This technique is a non-parametric estimation of a regression model
of a signal using an orthogonal basis \cite{Donoho95}-\cite{Neumann97}
and works best for signals whose decomposition has a small amount of detail
coefficients considerably different
from zero.

The reconstruction of a function $x(n)$ is carried out using the modified wavelet coefficients for the signal deconvolution:
\begin{equation}\label{Eq5}
\hat{x}(n) = \sum\limits_{m = 0}^{N_{j_o }  - 1}
{\hat{a}_{j_0,m} } \varphi _{j_0,m} (n) + \sum\limits_{j = j_0 }^J {\sum\limits_{m = 0}^{N_j  - 1} {\hat{d}_{j,m} \psi _{j,m} } } (n),
\end{equation}
where $\hat{a}_{j_0,m}$- and $\hat{d}_{j,m}$-coefficients passed wavelet thresholding.

As a rule, the criteria of entropy minimum $(H)$ is used to choose the optimum wavelet decomposition,
$Í$  being a logarithm of a signal's energy $H = \sum {log} \left( {\hat{x}^2 } \right).$

Entropy of an initial signal reaches its maximum due to noisiness of a signal. With the increase of the wavelet decomposition
the entropy decreases to its minimum, which corresponds to the optimal level of wavelet decomposition of the initial signal.

\subsection{
 Determination of the optimal noise reduction for a signal from LPC}
 When studying the effects constituting a small fraction of the background one should, apart from high quality of experimental data, impose high requirements for their processing. Setting of the input parameters for the arbitrary noisy signals can present some difficulties in our case due to the necessity of having a priori data or an analysis of statistical characteristics of high frequency components of the signals.
 Since in the typical double-beta decay measurements
 (thousands of hours) the spectrum of noise can be exposed to substantial changes due to possible electronic noise pickups and microphonic noise, some adaptive method of noise reduction is needed in order to minimize the uncertainty of the initial signal shape evaluation which substantially influences the determination of the type of event.

 It should be remarked that the choice of both a wavelet to be used and a level of decomposition depends on the properties of the signal under investigation. 
 More smooth the wavelets create more smooth approximation of a signal, and vice versa, "short" the wavelets trace peaks of approximated function is better.
 The level of decomposition affects the scale of sifted out details. With the increase of the level of decomposition the model subtracts the increasing level of the noise, with the result that possible soothing occurs not only of the noise but also of some local characteristics of a signal. A series of approximation and detail wavelet coefficients is needed for the wavelet transform.

The noisy component of a signal in most cases is reflected in the detail coefficients $\tilde{d}_{j,m}$, and it is these coefficients that are subject to processing in the noise reduction technique. The noisy component, as a rule, has an absolute amplitude value less than that of the basic signal. So to reduce the noise one should zero those coefficients that are less than some threshold value. The choice of the threshold level of noise  affects the quality of the signal noise reduction, which could be evaluated as a signal to noise ratio $(P)$:
\begin{equation}\label{Eq6}
P=\frac{{\frac{1}{{N_s }}\sum\limits_{n = 1}^{N_s } {s^2 (n)} }}{{\frac{1}{{N_z }}\sum\limits_{n = 1}^{N_z} {\left( {z(n) - \overline {z(n)} } \right)^2 } }},
\end{equation}
where $s(n)$ and $z(n)$ are the discrete vectors of a signal and noise; while $N_s$ and $N_z$ are their lengths, respectively.

Setting a small threshold keeps a residual noise in the detail coefficients and results only in insignificant increase in the signal to noise ratio.
With large enough threshold one can lose coefficients carrying essential information. Search for an optimum threshold amounts to finding the
maximum value of the signal to noise ratio with the least quadratic deviation of the estimated signal.

Let us express the empirical detail coefficients $\{{\tilde{d}_{j,m} } | m = \overline {1,N_j }  \}$
corresponding to level $j$ by the linear relation
\begin{equation}\label{Eq7}
\tilde{d}_{j,m}  = g_{j,m}  + \sigma _j \vartheta _{j,m},
\end{equation}
where $ \{ g_{j,m}  | m = \overline {1,N_j}  \}$ are the true detail coefficients of a signal without noise $(N_j$  in this case is the
number of detail coefficients, considered at a decomposition level $j)$, and
$\{ \vartheta_{j,m} | \vartheta _{j,m}  \in {\rm{Norm}} ({\rm{0}},\sigma _j^{2}){\rm{,}} m = \overline {1,N_j } \}$ are the reading of the additive Gaussian noise with zero mean and variance, $\sigma _j^{2}$.
Then the solution of the noise reduction problem could be reduced to the search of estimates $\{\hat{g}_{j,m} |m=\overline{1,N_j}  \}$ of true detail coefficients:  $g_{j,m}=T(\tilde{d}_{j,m})$. Such an estimate, carried out on the basis of empirical coefficients $\{\tilde{d}_{j,m} |m=\overline{1,N_j}\}$ and a given threshold $\theta _{j,i} $, is in fact a construction of a regression model of the true coefficients:
\begin{equation}\label{Eq8}
\hat{g}_{j,m}  = d_{j,m}  + \phi ( {d_{j,m} ,\theta _{j,i} } ),
\end{equation}
where $\phi ( {d_{j,m} ,\theta_{j,i} } )$ is the remainder term of thresholding function written in a general form. Expressing by the least-squares procedure the deviation of coefficients in the regression model (\ref{Eq8}) as some risk function
\begin{equation}\label{Eq9}
R_j({\theta _{j,i}}) = \sum\limits_{k = 1}^{K_j } {\mathop {({\hat{g}_{j,k}  - g_{j,k} })}\nolimits^2 },
\end{equation}
where $i \in N$, we learn that the optimum value of the threshold  $\overline \theta _j $, producing in accordance with the Stein criterion \cite{Stein81} the best noise reduction of a signal, corresponds to the case where function (\ref{Eq9}) has a global extremum:
\begin{equation}\label{Eq10}
\overline \theta  _j  =  \arg {}{}{} \min \limits_{\overline \theta_{j,i} | i \in  N} R_j ( {\theta_{j,i} } ).
\end{equation}

To meet this criterion used in the noise reduction we applied the Birg\'e-Massart technique \cite{Massart97} to determine the threshold $\overline \theta  _j $ to treat the detail wavelet coefficients. Coefficients $\tilde{d}_{j,m}$ less than the chosen threshold, were zeroed while the others were diminished by $\overline \theta_j $. The optimum value of criterion $\overline \theta_j $ has been chosen in accordance with minimum "entropy - logarithm of signal energy" principle.

To enhance the reliability of both empirical signal shape determination and its parameters' extraction we can add to each initial frame of
a noisy pulse a known model signal with preset parameters and then process it with wavelet analysis.  As a rule, during first $\sim 40$ $\mu s$ ("prehistory") in the frame of digitized pulse from the CSA [see Fig.~\ref{fig:real_pulse} $(c)$ and $(d)$], there is a noise path only.  In this interval of time we can select a vector of values containing no signal. Thus, we can produce a sampling control
$\{\tilde{z}(n)|n = 1,...,N_x/4;n = 1,...,N_x/4\}$ composed of two halves, each one being
equal to "prehistory" ($N_x$ is the total length of the frame). We add a model signal $s_o(n)$, composed of three Guassian curves summed together.
We derive a model signal with a real noise having put in linear dependence of a width and center of the Gaussian curves on the rise time of CSA output pulse,
\begin{equation}\label{noise s}
\tilde{s}(n)=s_o(n)+\tilde{z}(n), n = 1,...,N_x/2.
\end{equation}

To find the optimum technique for noise reduction of a similar signal and to define a quantitative measure of concordance between the estimated $\hat{s}(n)$ and initial $s_o(n)$ signals, we introduce the following criteria:
\begin{equation}\label{Eq11}
A=\frac{\sum\limits_{n = 1}^N [\hat{s}(n)-s_o(n)]^2}{\sum\limits_{n = 1}^N s_o^2(n)};
\begin{array}{*{20}c}
   {} & {}  \\
\end{array}
B=\frac{\sum\limits_{n = 1}^N \hat{s}(n)s_o(n)}{\sum\limits_{n = 1}^N s_o^2(n)};
\end{equation}
\begin{equation}\label{Eq13}
    C=\sqrt {A^2  + (B - 1)^2 }.
\end{equation}

The minimum value of functional $A$ corresponds to the identity of the estimated and
modeling signals, and its maximum corresponds to their total anticorrelation. In contrast,
the minimum value of functional $B$ indicates the anticorrelation of the mentioned signals
and its maximum standing for their total identity. So, the lesser $A$ and greater $B$, the
more accurate is the resurrection of a signal under investigation. In other words, the algorithm
of the signal $s_o(n)$  reconstruction amounts to minimization of functional $A(\hat{s})$
with simultaneous maximization of functional $B(\hat{s})$. The normalization of the introduced
functionals allows comparing of various signals' decomposition for different wavelet function basis irrespective of the parameters of the signals and  basic functions. Functional
$C(\hat{s})$ is a nonlinear combination of functional $A(\hat{s} )$ and functional $B(\hat{s})$, and serves as a generalized quantitative characteristic of a quality of the reconstructed signal and is to be minimized.

\subsection{Detection of a signal against additive noises}
The current task can be solved with the MATLAB using
Wavelet Toolbox \cite{MTBX} for signal and image treatment.
To choose the optimal wavelet allowing one to obtain the best of wavelet threshold of a signal, the criterion of maximum ratio of initial entropy
$(H_0)$ and that purified from the noise signal $(H)$ was used:
\begin{equation}\label{eq14}
\eta  = {{H_{\rm{o}} } \mathord{\left/
 {\vphantom {{H_{\rm{o}} } H}} \right.
 \kern-\nulldelimiterspace} H}.
\end{equation}

Confusing, you have to explain "tree" each type of wavelets using the criterion "entropy - logarithm of energy". The following orthogonal wavelets with compact carrier were used - Dobeshies ($db$N), Simlet ($sym$N) and Coiflet ($coif$N), where  N is the index number, designating the number of non-zero coeffeicients in filters. Wavelets are not symmetrical and are not periodic enough. Simlet wavelets are close to symmetric to certain degree.

\begin{figure}[pt]
\includegraphics*[width=3.15in,angle=0.]{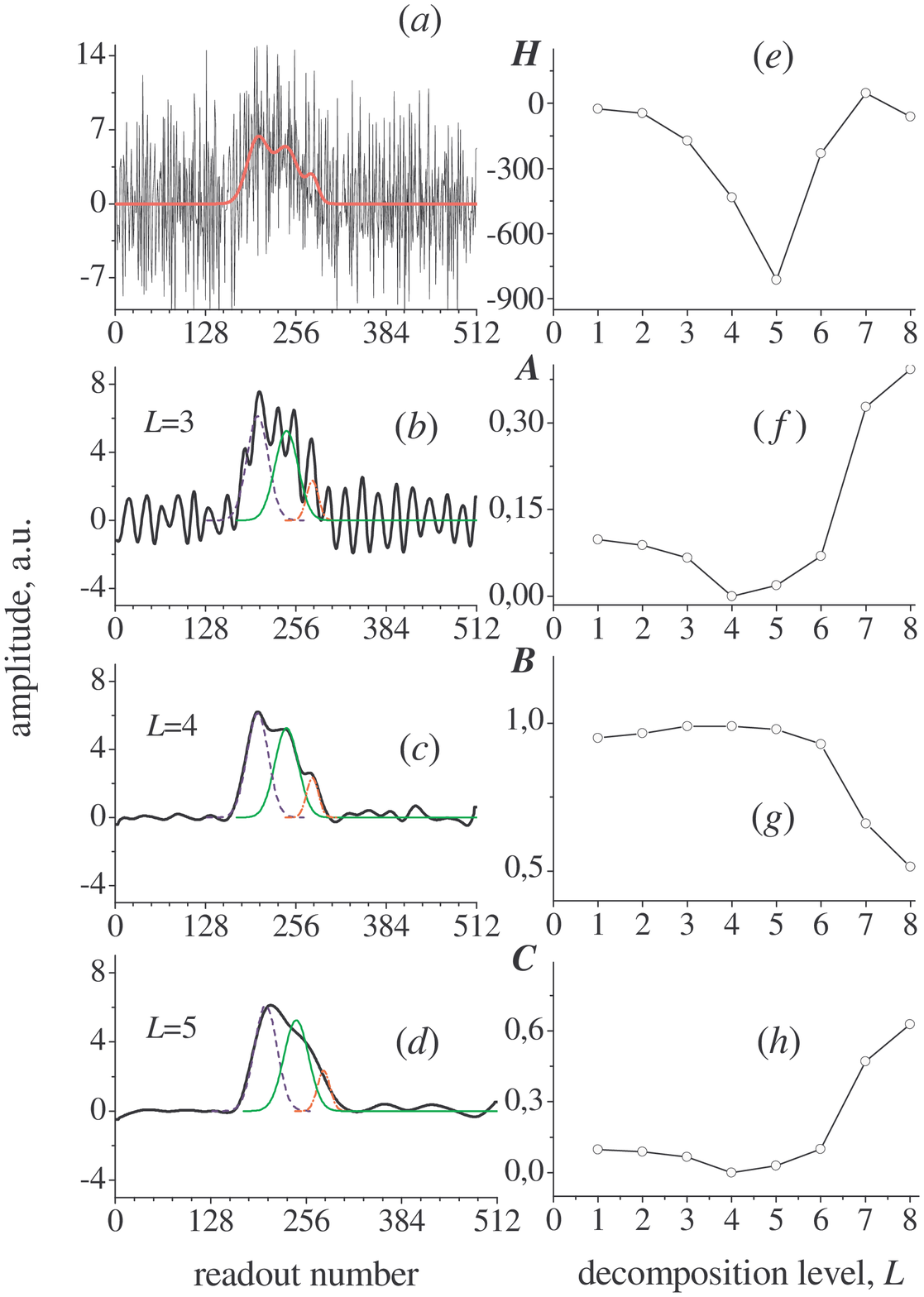}%
\caption{\label{fig:model and noise_pulse}
Signal modeling graphs
with the added real noise $(a)$ and de-noise of signals for various levels of decomposition: $(b)$-$L=3$, $(c)$-$L=4$ and $(d)$-$L=5$. $(e)$ and $(f)$, $(g)$, $(h)$ dependency's of entropy $H$ and coefficients  $A$, $B$, $C$ from wavelet decomposition level, respectively.}
\end{figure}

For each specified empirical signal we defined the optimal number of decomposition levels using numerical simulation for the case when the signal is resurrected to a large degree and the noise interference is not so large yet as to significantly distort the shape of a signal. The procedure of the signal resurrection is shown in Fig.~\ref{fig:model and noise_pulse}
for the model signal $s_o$ which has additively added  noise interference $z_o$ taken from a real signal $\tilde{x}(n)$. As one would expect, both the quality of the estimated signal and noise-to-signal ratio depend on the degree of overlapping of the wavelet-spectra of a signal to be reconstructed and noisy signal.

After choosing the best wavelet for a given pulse using the entropy criterion, the functionals (\ref{Eq11}) and (\ref{Eq13}) for each level of wavelet decomposition have been calculated for different values of threshold criterion of noise purification.  Values of $H$, $A$, $B$ and $C$ are presented in Fig.~\ref{fig:model and noise_pulse} in ($e$), ($f$), ($g$) and ($h$), respectively, for 8 levels of pulse wavelet decomposition of $\tilde{s}(n)$ [Fig.~\ref{fig:model and noise_pulse}$(a)$] using wavelet of Simlet $sym8$ and soft thresholding \cite{Donoho92}.

Analysis of Fig.~\ref{fig:model and noise_pulse} ($b$), ($c$) and ($d$) allows one to observe the change in the estimated signal with increase in number of decomposition levels $L$ taken into account. Fig.~\ref{fig:model and noise_pulse}$(c)$ shows that for decomposition level  $L=4$ of a noise signal $\tilde{s}(n)$ the minimum of functional $C(\hat{s})$  corresponds to the maximum accuracy in model signal $s_o(n)$. With the decomposition level of $L=3$ we obtain the significant increase in sinusoidal noise, in case of $(L=5)$, which corresponds to the minimum of entropy $H$, the estimated signal takes smoother shape resulting in stronger distortions of the initial model signal parameters.

Information on the type of event is contained in the shape of a electric current signal.
Therefore the minimum of functional $C(\hat{s})$ approaches in our case the quantitative criterion describing a relation between an initial modelling signal without noise and estimated signal is better.

Thus, by determining, with the help of a model signal with definite empirical noise, the depth of the wavelet decomposition and threshold values $\overline{\theta}_j$ for each decomposition level we find the resurrected empirical signal proper to the event in LPC. By describing signals purified from noise with the help of a set of Gaussian curves using the minimization root-mean-square error technique one can separate multipoint events from single-point ones.

\section{Results of the first stage of measurements using LPC}
\begin{figure}[pt]
\includegraphics*[width=3.05in,angle=0.]{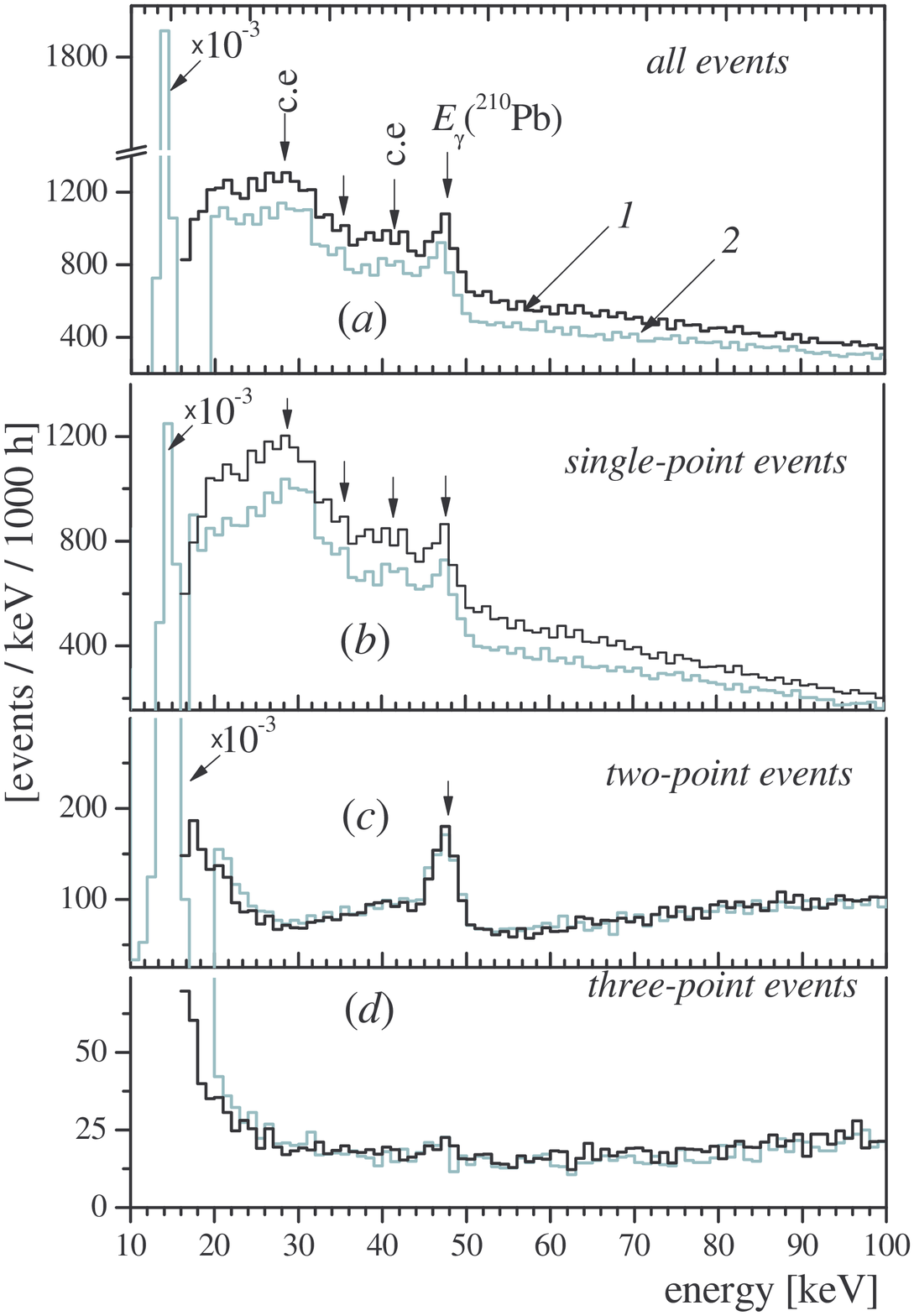}%
\caption{\label{fig:spc_run} Pulse height $M1$ spectrum's (normalized for 1000 h) of the background of LPC filled with krypton enriched in $^{78}$Kr ("{\it 1}", dark line)
and with natural krypton ("{\it 2}", light line): $(a)$ - all events,  $(b)$ - single-point events, $(c)$ - two-point events and $(d)$ - three-point events.}
\end{figure}

Fig.~\ref{fig:spc_run}$(a)$ shows the total
spectra of $M1$ amplitudes,  normalized for 1000 h,  obtained for the background of LPC filled with krypton enriched in $^{78}$Kr (total collection time is 8400 h; spectrum "{\it 1}", dark line), and with natural krypton (total collection time is 5000 ÷; spectrum "{\it 2}", light line).
Comparison of these spectra shows that the background of the counter filled with krypton enriched in $^{78}$Kr exceeds the background with natural
krypton at energy $\geq18$ keV.
The shape of difference (subtracted) spectrum
in the range of $20-100$ keV is
described reasonably well by the model spectrum
composed of $\beta$-spectra due to $^{85}$Kr and $^{14}$C decay in the proportion 1:17. Background counting rates in this range are
56.3 h$^{-1}$ and 47.2 h$^{-1}$ when filled with $^{78}$Kr and $^{nat}$Kr, respectively.
The difference in $^{85}$Kr isotope activities in the samples is due to its different residual value achieved during the isotopic purification. $^{14}$C isotope comes into the krypton gas during the continuous usage of the counter, supposedly as a result of slow sublimation of organic molecules (ethyl alcohol and acetone were used  to clean the components when assembling the detector) from the surface of the body of the counter. Overactivity of $^{14}$C in the enriched krypton is due to the fact that this sample was measured first and the main portion of this organic vapour penetrated into it. Actually, all the events in the operating energy range produced by $\beta$-particles absorption can be considered as single-point ones, except for those event where $\beta$-particles have lost partially their energy through generating bremsstrahlung photon.

At energies $\leq18$ keV, the counting rate of events in spectrum "\emph{2}" is much higher than in
spectrum "\emph{1}". This is due to the presence, in the original atmospheric krypton, of
cosmogeneous radioactive isotope $^{81}$Kr ($\mathrm{T}_{1/2}=2.29\cdot10^5$ yrs) \cite{nudat2}  with volume
activity of $\sim 0.1$ min$^{-1}l^{-1}$Kr \cite{Loosli68}-\cite{Kuz80}; the significant part of which comes into the sample of natural krypton during the process of its production. $^{81}$Kr decays by
electronic capture producing $^{81}$Br (\emph{K}-capture 87.5\% \cite{Chew74}). \emph{K}-capture yields a 13.5 keV energy release. One can see the peak of this line in spectrum "\emph{2}" (light line in Fig.~\ref{fig:spc_run}). The resolution of the peak (FWHM) is 15.3\%. Counting rate of events in $(13.5\pm3.0)$ keV energy range is 220 h$^{-1}$, which corresponds to the volume  activity of $^{81}$Kr $(0.10\pm0.01)$ min$^{-1}l^{-1}$Kr.

The filling of \emph{K}-shell vacancy of a daughter atom of bromine in 61.4\%  goes with emittance of
characteristic X-rays of 11.92 keV ($K_{\alpha 1}$, 100\%), 11.88 keV ($K_{\alpha 2}$, 51.9\%),
13.29 keV ($K_{\beta 1}$, 13.6\%), 13.47 keV ($K_{\beta 2}$, 1.36\%)
energies \cite{nudat2} and accompanying them Auger electrons of 1.55 keV, 1.60 keV,
0.27 keV and 0.01 keV, respectively (relative intensities of $K_{\alpha \beta}$-lines
are given in parentheses). Thus it is clear that in the case of photon
emittance $K_{\alpha 1}$, $K_{\alpha 2}$ energy of Auger electrons
is large enough to produce a distinct two-point event.
Thirty-eight point six
percent of all the events are due to the filling of Br atom \emph{K}-shell vacancy accompanied by a
cascade of Auger electrons, and are considered as events producing one point energy release.
Events with the emittance of characteristic X-rays $K_{\alpha 1}$ and $K_{\alpha 2}$ are also
considered as one point events since energy release of Auger electrons do not exceed the
CSA noise. Using the above data and taking into account the efficiency of characteristic
X-rays absorption in a krypton gas of LPC $(\varepsilon _p=0.869)$ one can calculate the
component analysis of complete absorption peak with energy 13.5 keV with 49.4\% single-point
+ $50.6$\% two-point event \cite{PTE}.

At 46.5 keV, in spectra "\emph{1}" and "\emph{2}" [Fig.~\ref{fig:spc_run}$(a)$], one can see the peak corresponding to the source
of $^{210}$Pb (T$_{1/2}=22.2$ yrs, $\beta$-decay, $E_\gamma=46.5$ keV, with a yield is 4.25\% per decay \cite{nudat2}). $^{210}$Pb istope
is produced after $^{222}$Rn (T$_{1/2}=3.82$ days, $\alpha$-decay) in the $^{238}$U decay chain. It can be produced directly in copper
material due to trace radioactive contaminant decay (volume source) or accumulate on the surface of the counter body in the form
of daughter isotopes of $^{222}$Rn decay, $^{222}$Rn being present in the atmospheric air (surface source),
during the preparation of mounting the detector.
Gamma-radiation comes into a
fiducial volume
of the counter from both sources. There are mainly single-point events (due to photoeffect on krypton with de-excitation due to Auger-electrons) and two-point events (due to photoeffect with emission of krypton characteristic radiation) in the peak line of 46.5 keV, and only a small portion of events will be regarded as three-point ones where the primary photon or characteristic photon absorption occurred through the scattering on the outer electrons with the following absorption of the secondary quantum.

Conversion electrons
(c.e.) can originate from the surface
source, $^{210}$Pb.
30.1 keV (52\% per decay) and 43.3 keV (13.6\% per decay) lines are the most intensive.
There are peaks in spectra \emph{1} and \emph{2} at $\sim 28$ keV and $\sim 41$ keV
that could be associated with these lines. All the events of these peaks  should be single-point ones. The observed shift of maximum energy
with respect to the expected line could be explained by the deposit of a portion of near-wall primary ionization electrons on the cathode
due to their diffusion.
With the emittance of conversion electrons the residual excitation
of the $^{210}$Bi daughter shell deenergizes by radiation of
characteristic photons of $L$-series ($E_{L_\alpha}=10.8$ keV,
9.3\% per decay; $E_{L_\beta}=13.0$ keV, 11.2\% per decay and etc.)
or/and by Auger electrons. In different combinations this
radiation can enter the fiducial volume simultaneously with
$\beta$-particles and c.e. In the simplest case, when c.e.
with 30.2 keV and  characteristic $L_\alpha$-photon are registered,
a two point event with energy deposit of $\sim42.5$ keV is
produced in the gas.  The fraction of such events is small enough
in the total spectrum. There is a step at $\sim34$ keV
[Fig.~\ref{fig:spc_run} $(a)$ and $(b)$]
corresponding to the escape peak of 46.5 keV line for krypton
($E_\gamma-E_{K_{\alpha \mathrm{Kr}}}=46.5-12.6=33.9$ keV).

In general, all types of distortion of spectrum \emph{1}, at the energies lower than $\sim50$ keV, from  the smooth drop-down base which is characteristic for energies higher than 50 keV, could be "assigned" to $^{210}$Pb source.

The scenario described above is confirmed
by the distribution of events related to different peaks in the amplitude spectra, composed separately for events with different number of point ionization clusters in the composite pulse.  In Fig.~\ref{fig:spc_run} one can see the corresponding spectra of amplitude $M1$ plotted after selecting the type of events by analyzing  the form  of electric current pulses purified from noise: ($b$) - single-point, ($c$) - two-point, ($g$) - three-point.

Comparison of the obtained relation of peak areas (13.5 keV and 46.5 keV) for different components with estimated values allow one to determine the efficiency, $\varepsilon_k$, of the procedure of the computer event selection with a given number of ionization regions. The efficiencies, $\varepsilon_k$, of two-point event selection were found to be 0.567 and 0.733 respectively, which are in good agreement with calculated model values - 0.604 and  0.745 \cite{PTE}.

As seen from Fig.~\ref{fig:spc_run}$(b)$, the main difference
for background spectra for LPC filled with different gases
is in single-point (event) spectra. LPC background spectra with various gases are in good agreement with each other for two- and three-point events in the energy region of $20-80$ keV. For further analysis the three-point event spectra, Fig.~\ref{fig:spc_run}$(d)$, have been used.
Each event of these spectra is characterized by a set of energy deposits distributed over three point-like regions of the operating volume
of the counter. Energy deposits are proportional to the Gaussian areas ($A_1$, $A_2$, $A_3$), which describe three pointwise components
of the total/composite current pulse of primary ionization electrons on the border of the gas amplification region.
Gaussian numbering corresponds to the arrival time sequence of the components coming into the gas amplification region.
To simplify the selection of events with a given set of features the amplitudes of components for each event  are arranged in the
increased order $[(A_1, A_2, A3) \rightarrow (q_0 \leq q_1 \leq q_2)]$.

In the sought-for events the minimal amplitude value
(two groups of Auger-electrons from residual excitation with energies $E_1 \sim 1.5$ keV and  $E_2 \sim2.9$ keV) will correspond
to a larger extent with respect to the resolution to the condition  0.9 keV $\leq q_0 \leq 4.5$ keV: "\emph{C1}". The middle and highest
amplitudes are produced by characteristic \emph{K}-photons
$(K_{\alpha1}+ K_{\alpha1},K_{\alpha1}+ K_{\alpha2},K_{\alpha2}+
K_{\alpha2},K_{\alpha1}+ K_{\beta1},K_{\alpha1}+ K_{\beta2},K_{\alpha2}+ K_{\beta1},K_{\alpha2}+ K_{\beta2})$ Se$^{**}$.
In the total number of possible combinations, the fraction of the enumerated combinations of the two photon $(\alpha_k)$,
registered in a composition of three-point event is $\alpha_k=0.985$.
It's convenient to introduce a parameter of average to maximum amplitude ratio. With an allowance for resolution
significant part of the event
would have this parameter within $1.0\geq q_1/q_2\geq0.7$: "\emph{C2}". The selection of events from spectra in Fig.~\ref{fig:spc_run}$(d)$ corresponding to the conditions ("\emph{C1}"+"\emph{C2}") allows an additional decrease in the background of the expected peak energy region and gives a general representation of a spectra form for wide range of energies.

\begin{figure}[pt]
\includegraphics*[width=3.20 in,angle=0.]{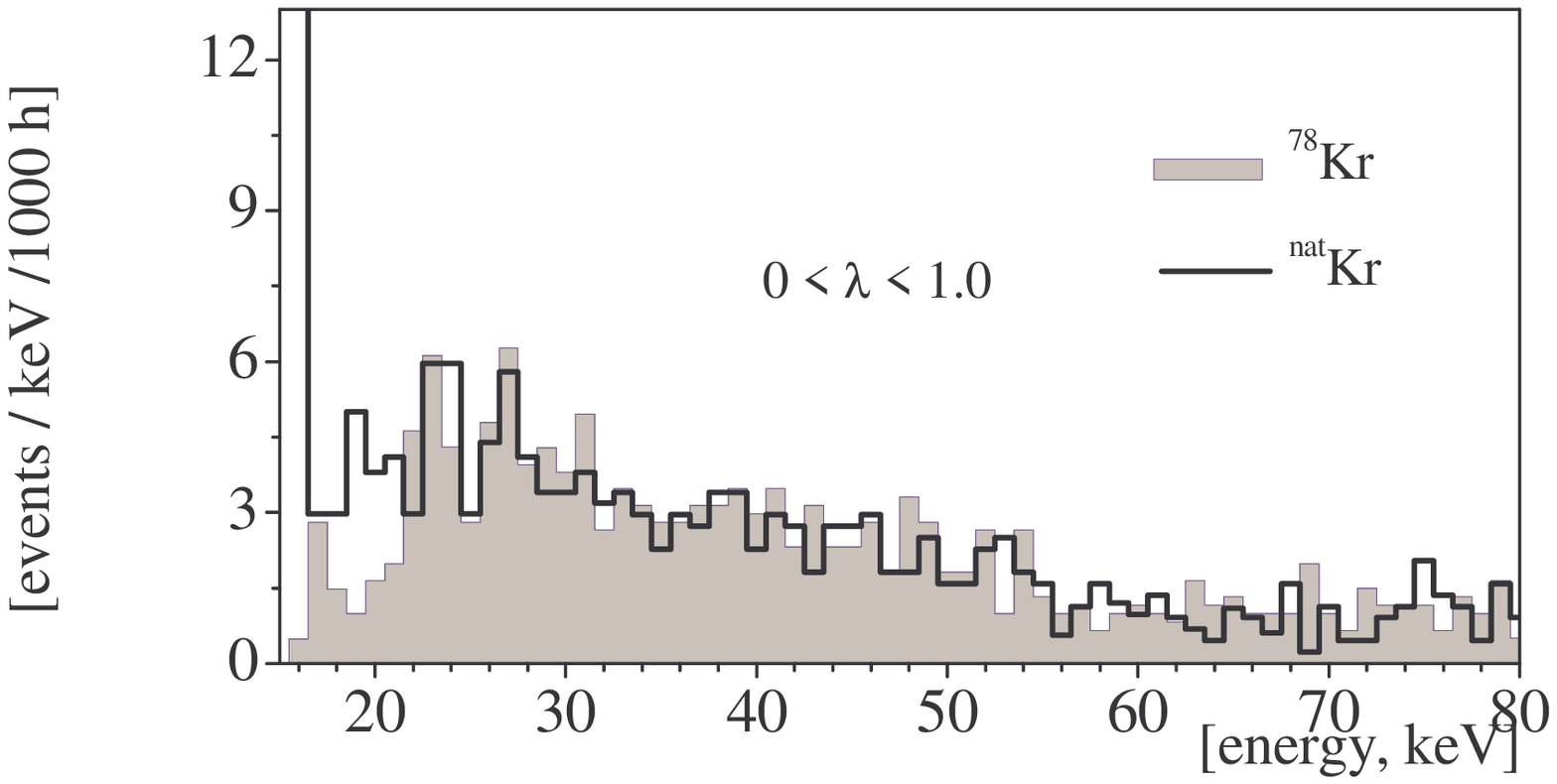}%
\caption{\label{fig:spc_select run}
The three-point spectra selected under the conditions "\emph{C1}" and "\emph{C2}".}
\end{figure}

\begin{figure}[pt]
\includegraphics*[width=2.75in,angle=0.]{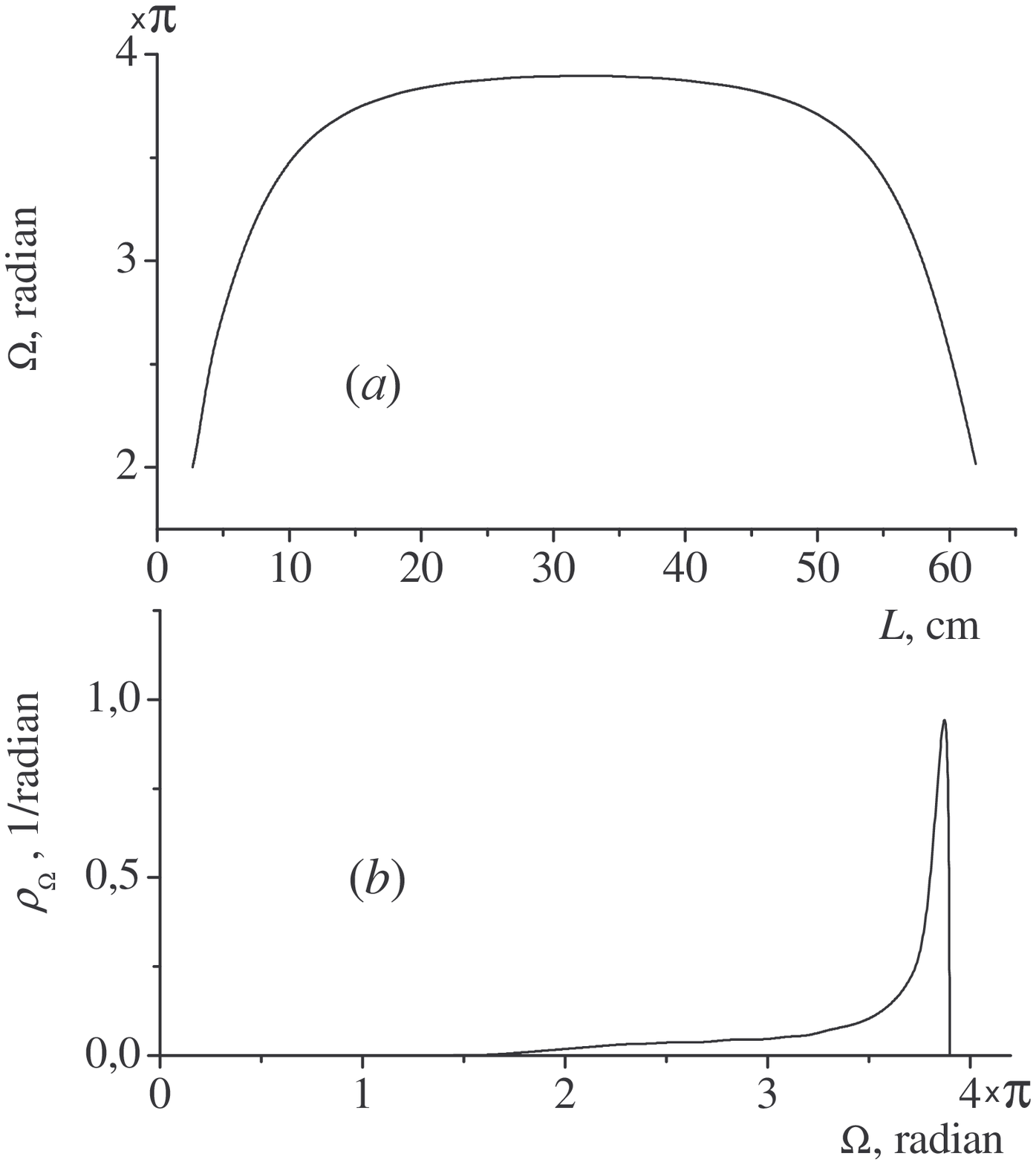}%
\caption{\label{fig:solid angle} $(a)$ - the solid angle under which the interior surface of the counter from different points of an anode wire is visible.
$(b)$ - the density of distribution of the solid angle for the points uniformly distributed along the anode wire.}
\end{figure}

The selected spectra are plotted in Fig.~\ref{fig:spc_select run}. They show that
there are similar peaks for both spectra in the energy region of $\sim26$ keV corresponding to the sought-for effect. To find their nature the study of the distribution of the number of background events of various types along the anode wire has been performed.
This distribution was plotted using the dependence of the relative amplitude of the first after-pulse  on the distance between the point of origin of the main pulse and the middle point of the length of the anode wire. This distance defines the solid angle viewing the inner surface of the copper cathode cylinder. The solid angle through which the operating surface of the cathode is viewed from the middle point of the anode wire and from the end points of the operating length of the anode wire are $\sim 3.9\pi$ and $\sim 2\pi$, respectively [Fig.~\ref{fig:solid angle}$(a)$]. The density of distribution of the solid angle for the points uniformly distributed along the anode wire is presented in Fig.~\ref{fig:solid angle}$(b)$.
There is a dependence on the solid angle of the relative number of photoelectrons, knocked out of the copper surface by
the photons produced in the gas ionization.
First after-pulse is produced as a consequence of gas amplification of secondary photoelectrons generated on the cathode within the operating length of the anode wire.
It is the $\lambda$ parameter, equal to the ratio of after-pulse and pulse amplitudes,
$(M2-M1)/M1$, accurate to the precision set by energy resolution of pulse and after-pulse, that determines the coordinate of the primary event with
respect to the middle point of the anode wire. To calibrate the counter with respect to $\lambda$ one needs a radioactive source
uniformly distributed over the LPC volume.
$^{81}$Kr is well suited for this task.

\begin{figure}[pt]
\includegraphics*[width=2.75in,angle=0.]{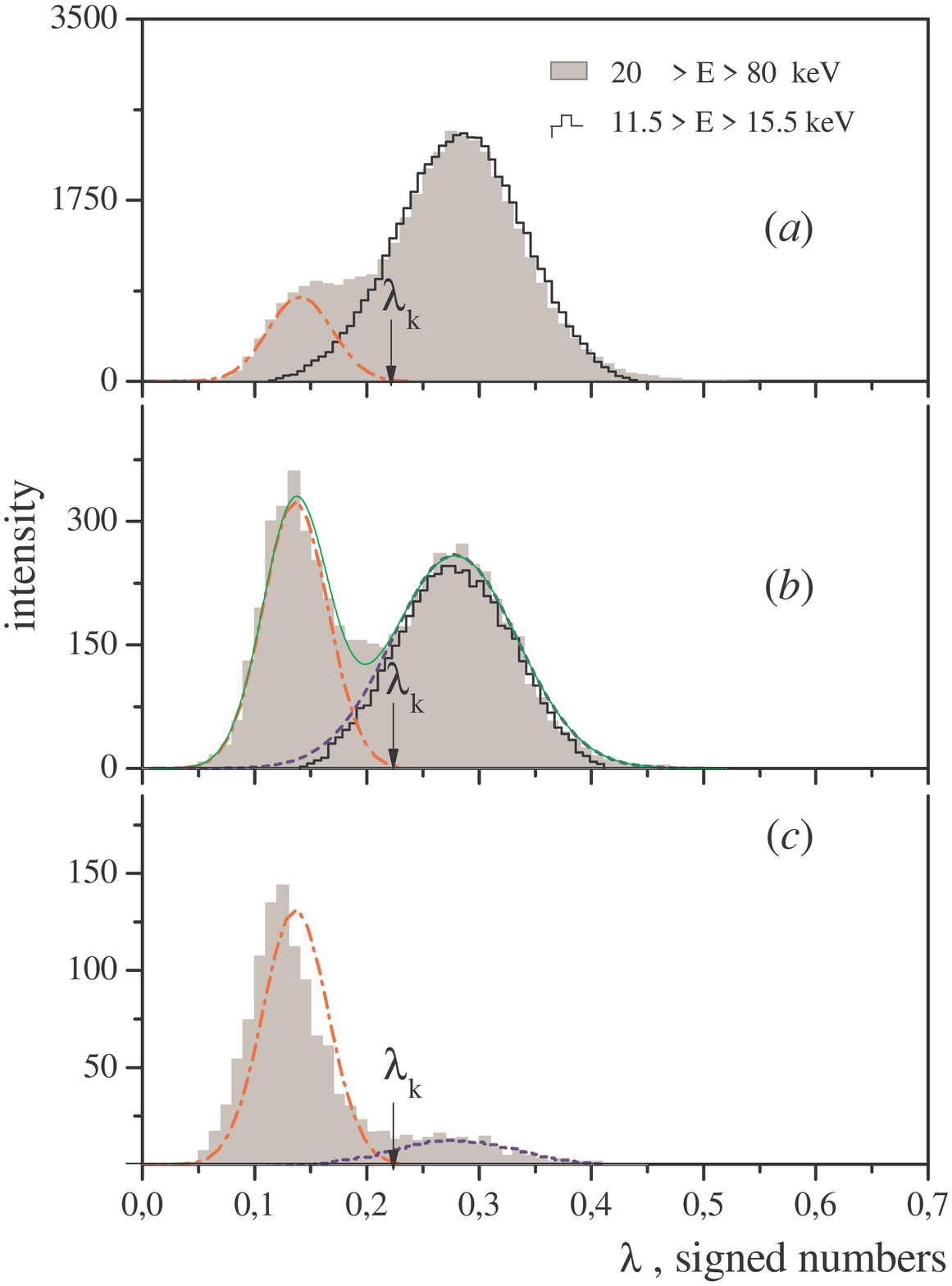}%
\caption{\label{fig:spc lambda} The distributions of the events in LPC filled with natural krypton with energies $11.5-15.5$ keV ($^{81}$Kr, histogram) and with energies $20-80$ keV (shaded region). The dash-dot and dotted curve is result of fitting operation of the shaded region. $(a)$ - single-point events, $(b)$ - two-point events and $(c)$ - three-point events.}
\end{figure}

A normalized distribution of the number of events of $11.5-15.5$ keV energy ($^{81}$Kr) from the spectrum of one point events of the LPC background with natural krypton versus parameter $\lambda$ is given in Fig.~\ref{fig:spc lambda}$(a)$ as histogram with its maximum at $\lambda=0.28$.
Distribution of the number of event with $20-80$ keV energy range is given in the same figure in shaded region.

Comparison of the distributions shows an excess of events over a bar graph in the region of $\lambda < 0.225$ value, corresponding to the ends of the anode wire.
The form of distribution of this excess is obtained by fitting the subtracted out histogram ["shaded region" minus "histogram"], and is
presented by dash-dot curve.  Hence it follows that at
the ends of the operating length of the anode
wire there is an extra background source additional to the one that is uniformly distributed over the volume. In Fig.~\ref{fig:spc lambda} $(b)$ and $(c)$, there is similar distribution of the number of two-point and three-point events with energy $20-80$ keV (shaded region) and their uniform and near-boundary components (dotted and dash-dot curves). It is clear that the relative contribution of near-boundary into the corresponding spectrum increases with the increase of the number of point-like ionization clusters in a given event.

\begin{figure}[pt]
\includegraphics*[width=2.60 in,angle=270.]{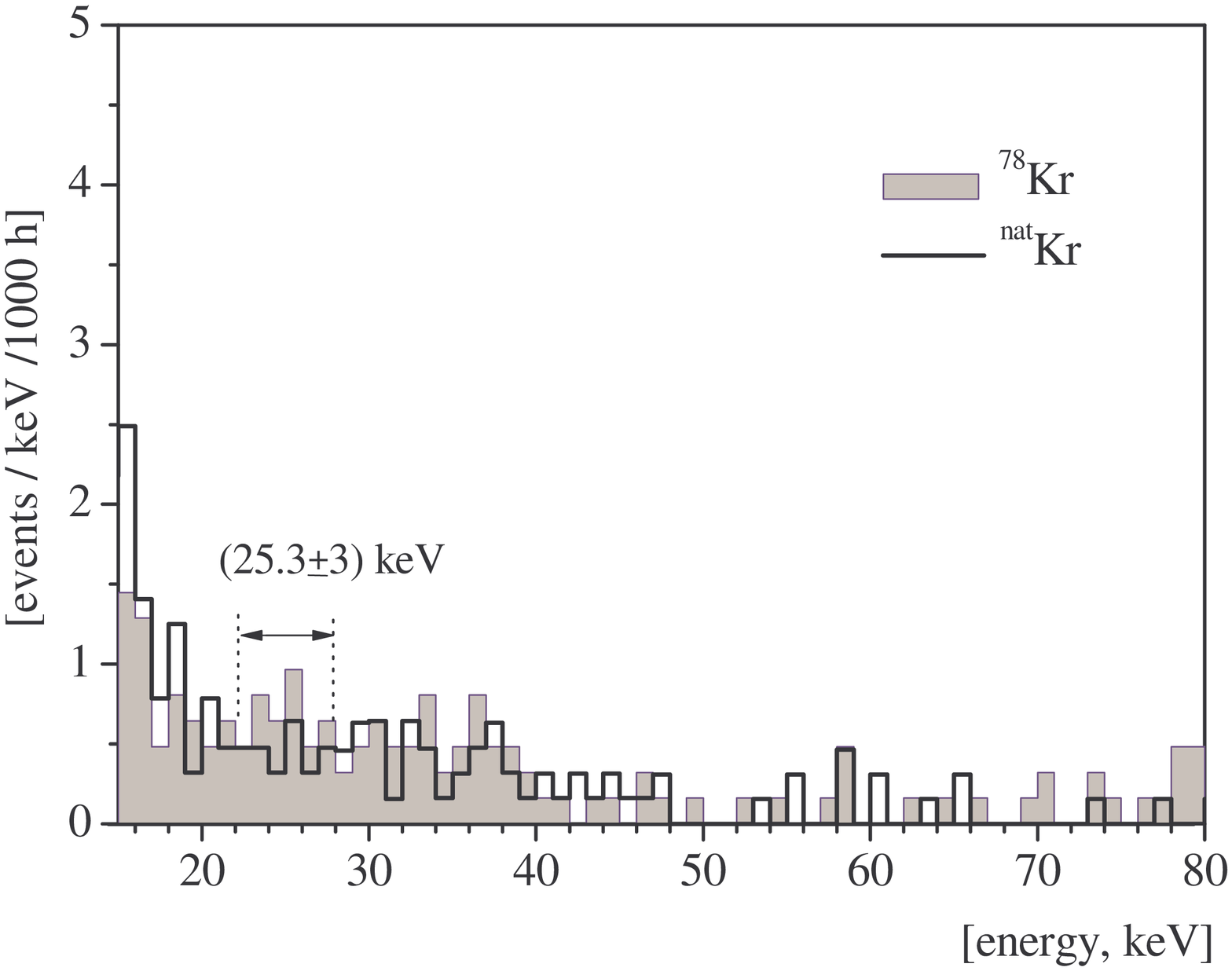}%
\caption{\label{fig:res spc3}
The three-point spectra selected under the conditions ("\emph{C1}"+"\emph{C2}") and $\lambda \geq 0.225$.}
\end{figure}

The additional events at the ends of the operating length of the anode wire are produced by particles coming out of the 'dead' near-end volumes of a krypton gas. Apart from the proportionally amplified components a total pulse could also be composed of components of larger energy deposits that occur in the 'dead' volume and are collected in the ionization mode at the end bulbs of the anode. Specifically, the $\sim26$ keV in spectra of Fig.~\ref{fig:res spc3}$(a)$
could well be formed by alpha-particles emitted from the surface of the cathode in the near-end region where the anode wire is thickened with copper tubules.
Alpha-particle can ionize \emph{K}-shells of krypton atoms, and in this case the characteristic photons would have energies of 12.6 keV.
These photons can penetrate into the
operating volume of LPC where gas amplification occurs. Registration of two photons gives two points. The third point is produced by ionization of alpha-particle collected in the ionization mode. Discarding pulses with $\lambda < 0.225$ one can completely eliminate such events from 'three-point' spectra. The selection coefficient of useful event ($k_\lambda$ ) remains high enough: $k_\lambda=0.84$.

In Fig.~\ref{fig:res spc3} shaded-to-zero the bar graph ($^{78}$Kr) and the linear bar
($^{nat}$Kr) are three-point spectra selected under condition of $\lambda \geq 0.225$ from spectra shown in Fig.~\ref{fig:spc_select run}. The background in energy region of $(25.3\pm3.0)$ keV in these spectra is $n_1=(38.1\pm6.3)$ yr$^{-1}$ and  $n_2=(20.7\pm6.0)$ yr$^{-1}$, respectively, which gives count rate of $2K$-capture for $^{78}$Kr to be $n_{exp}=n_1-n_2=(17.4\pm8.7)$  yr$^{-1}$. The obtained positive value does not exceed two standard deviations and in this case limit on the expected effect only was set for 90\% C.L.: $n_{exp} \leq 31.7$ yr$^{-1}$.

The half-life limit has been calculated using formula
\begin{equation*}
\mathrm{T}_{1/2} \geq (ln2) \cdot N \cdot p_3 \cdot \varepsilon_p \cdot
\varepsilon_3 \cdot \alpha_k \cdot k_\lambda  /  n_{exp},
\end{equation*}
where $N=1.08\cdot10^{24}$ is the number of $^{78}$Kr atoms in
the operating volume of the counter, $p_3=0.355$ is a portion of $2K$-captures accompanied by the emission of two \emph{K} X-rays;
$\varepsilon_p=0.809$ is the probability of two \emph{K}-photon
absorption in the operating volume; $\varepsilon_3=0.422$ is the selection efficiency for three-point events due to $2K$-capture in $^{78}$Kr;
$\alpha_k=0.985$ is the portion of events with two \emph{K}-photon that
could be registered as distinct three-point events;
$k_\lambda=0.840$ is the useful event selection
coefficient for a given threshold with respect to $\lambda$.
The result obtained is
\begin{equation*}
    \texttt{T}_{1/2} (0\nu+2\nu,2K) \geq 2.4\cdot 10^{21}\texttt{yr}~(90\%~\texttt{C.L.})
\end{equation*}

\section{CONCLUSION}
We have demonstrated the possibilities for reducing the background of the proportional counter by $\sim2000$ times for the case of
registering $2K$-capture events in $^{78}$Kr by selecting pulses according to the number of point-like clusters and the event's
coordinate along the length of the anode wire. In the spectrum of selected three-point events for LPC filled with enriched krypton, for the energy region of the sought-for
effect, there is an excess of events that, within the achieved level of statistics,
does not exceed two standard deviations.
The current result has been used to set the limit on the half-life of $^{78}$Kr $2K$-capture.
The experiment is in progress.

\begin{center}
    {\textbf{ACKNOWLEDGMENTS}}
\end{center}

We are very thankful to Prof. V.N.~Gavrin for allocation of  the experimental equipment in the underground laboratory of the Gallium Germanium Neutrino Telescope and also for his contribution in the early stages of this work.
The authors is grateful to Dr. Boris Pritychenko and  Dr. Timothy Johnson for productive discussions and careful reading of the manuscript and useful suggestions, respectively.
This work was supported by the Russian Foundation for Basic Research (Grant No. 04-02-16037)  and the Neutrino Physics Program of the Presidium of the Russian Academy of Sciences.


\begin{references}

\bibitem{Zdesenko} V.I.~Tretyak and Y.G.~Zdesenko, Atom. Data Nucl. Data Tables \textbf{80}, 83 (2002).
\bibitem{Barabash06} A.S.~Barabash, Czech.J.Phys.,  \textbf{56}, 437 (2006).
\bibitem{Danevich05} F.A.~Danevich {\it et al.}, NIM Phys.Res. \textbf{A544}, 553 (2005);
P.Belli {\it et al.}, Phys.Lett. \textbf{B658}, 193 (2008).
\bibitem{Barabash09} A.S.~Barabash {\it et al.}, Phys.Rev. \textbf{C80}, 035501 (2009)

\bibitem{Rukhadze09} N.I.~Rukhadze, Bull.Rus.Acad.Sci.Phys. \textbf{73}, 741 (2009).

\bibitem{r1} Ju.M.~Gavriljuk {\it et al.}, Phys. At. Nucl., {\bf69}, 2124 (2006).
\bibitem{r2}M. Aunola, J. Suchonen, Nucl. Phys. \textbf{A602}, 133 (1996).
\bibitem{r3}M. Hirsch, Nucl.Phys. \textbf{A577}, 411c (1994).
\bibitem{r4}O. Rumyantsev, M. Urin, Phys. Lett. {\bf B443}, 51 (1998).
\bibitem{r5}M. Doi, T. Kotani, Progr. Theor. Phys. v.{\bf{87}}, 1207 (1992).


\bibitem{x-ray} X-Ray-Data-Booklet, Center for X-ray Optics and Advanced Light Source LBNL, http://www.scribd.com/doc/; http://xdb.lbl.gov

\bibitem{Storm73}
E.Storm and H.I.Israel,
Atom. Data Nucl. Data Tables \textbf{7}, 565 (1970).

\bibitem{Drift and diffusion} A. Peisert and F. Sauli, {\it Drift and diffusion of electrons in gases: a compilation}, Preprint CERN 84-08, 13 July 1984.


\bibitem{r9}V.N.Gavrin {\it et al.},
Preprint INR RAN, P-698, Moscow (1991).

\bibitem{Kr-78purf}  A.N.~Shubin  {\it et al.}, Deep purification of krypton highly enriched in Kr-78 from Kr-85 with a gas
         centrifuge cascade, {\it Proc. of VII Inter. Scien. Conf. Physical and chemical processes on selection of atoms and molecules,
         Zvenigorod, October 6-10, 2003}. Moscow, Atominform; Troitsk, RSC RF TRINITI, 11, (2003).
\bibitem{PTE}Yu.M. Gavrilyuk {\it et al.}, Instr.Exper.Techn. {\bf 53}, No{\bf 1}, 57. (2010).
\bibitem{Daub92} I. Daubechies, Ten Lectures on Wavelets,  SIAM, (1992).
\bibitem{Wavelets} Wavelets in Physics, edited by J.C. van den Berg (Cambridge University Press,Cambridge,England,1999).
\bibitem{Mallat89} S. Mallat,IEEE Trans. Pat. Anal. Mach. Intell., \textbf{11}, 674, (1989).
\bibitem{Mallat99} S. Mallat, A wavelet tour of signal processing, New York: Academic, (1998).
\bibitem{Donoho95} D.L.~Donoho and I.M.~Johnstone, J.Amer.Stat.Assoc., \textbf{90}, No\textbf{432}, 1200, (1995).
\bibitem{Neumann97} M.~Neumann and R.~Sachs, Annals of Statistics, \textbf{25}, 38, (1997).
\bibitem{Stein81} C.M.~Stein,
Annals of Statistics, \textbf{9}, 1135,  (1981).
\bibitem{Massart97} L.~Birg\'e, P.~Massart, {\it From model selection to adaptive estimation}, in D. Pollard (ed), Festchrift for L. Le Cam, Springer, (1997).

\bibitem{MTBX} http://www.mathworks.com/products/wavelet/

\bibitem{nudat2} Evaluated Nuclear Structure Data File (ENSDF), http://www.nndc.bnl.gov/ensdf.

\bibitem{Loosli68} H.H.~Loosli, H.~Oeschger,  Earth Plan. Sci. Lett. \textbf{7}, No\textbf{1} 67, (1968).
\bibitem{Kuz80} V.V.~Kuzminov, A.A.~Pomansky,  Radiocarbon, \textbf{22}, No\textbf{2}, p.311, (1980).
\bibitem{Chew74} W.M.~Chew {\it et al.}, Nucl.Phys. \textbf{A229}, No\textbf{1}. 79,(1974).
\bibitem{Donoho92} D.L.~Donoho, {\it De-Noising via Soft Thresholding}, Tech. Rept., Statistics, Stanford, (1992).

\end{references}
\end{document}